\documentclass[preprint]{aastex62}
\accepted{April 8, 2019}
\shortauthors{Miao et al.}

\newcommand{\speed}[1]{#1 km~s${}^{-1}$}

\newcommand{\kms}{km~s$^{-1}$}
\newcommand{\degree}{\ensuremath{^\circ}}
\usepackage{longtable}
\usepackage{amssymb}
\usepackage{amsmath}
\usepackage{subfigure}
\usepackage{longtable}

\newcommand{\HeII}{\ion{He}{2}}

\newcommand{\rsun}[1]{${#1}\,R_\sun$}

\begin{document}

%\title{A NEW SMALL SATELLITE SUNSPOT TRIGGERING A RECURRENT JET EVENT ASSOCIATED WITH BLOWOUT JETS AND STANDARD JETS}
\title{A new small satellite sunspot triggering recurrent standard- and blowout-coronal jets}

\correspondingauthor{Yuhu Miao}
\email{myh@ynao.ac.cn}
\author[0000-0003-2183-2095]{Yuhu Miao}
\affiliation{Yunnan Observatories, Chinese Academy of Sciences, Kunming
650216, China}
\affiliation{Key Laboratory of Geospace Environment, Chinese Academy of Sciences, University of Science $\&$ Technology of China, Hefei 230026, China}
\affiliation{Department of Physics and Astronomy, King Saud University,
PO Box 2455, Riyadh 11451, Saudi Arabia}
\affiliation{Radio Cosmology Lab, Department of Physics, Faculty of Science,
University of Malaya, 50603 Kuala Lumpur, Malaysia.}
\affiliation{Institute of Space Physics, Luoyang Normal University, Luoyang, China}
\affiliation{University of Chinese Academy of Sciences, Beijing 100049, China}

\author{Y. Liu}
\affiliation{Yunnan Observatories, Chinese Academy of Sciences, Kunming
650216, China}
\affiliation{Key Laboratory of Geospace Environment, Chinese Academy of Sciences, University of Science $\&$ Technology of China, Hefei 230026, China}
\affiliation{Department of Physics and Astronomy, King Saud University,
PO Box 2455, Riyadh 11451, Saudi Arabia}
\affiliation{Radio Cosmology Lab, Department of Physics, Faculty of Science,
University of Malaya, 50603 Kuala Lumpur, Malaysia.}
\affiliation{Institute of Space Physics, Luoyang Normal University, Luoyang, China}

\author{Y. D. Shen}
\affiliation{Yunnan Observatories, Chinese Academy of Sciences, Kunming 650216, China}
\affiliation{Key Laboratory of Geospace Environment, Chinese Academy of Sciences, University of Science $\&$ Technology of China, Hefei 230026, China}
\author{A. ELMHAMDI}
\affiliation{Department of Physics and Astronomy, King Saud University,
PO Box 2455, Riyadh 11451, Saudi Arabia}

\author{A. S. KORDI}
\affiliation{Department of Physics and Astronomy, King Saud University,
PO Box 2455, Riyadh 11451, Saudi Arabia}

\author{H. B. Li}
\affiliation{Yunnan Observatories, Chinese Academy of Sciences, Kunming
650216, China}
\affiliation{University of Chinese Academy of Sciences, Beijing 100049, China}

\author{Z. Z. ABIDIN}
\affiliation{Radio Cosmology Lab, Department of Physics, Faculty of Science,
University of Malaya, 50603 Kuala Lumpur, Malaysia.}

\author{Z. J. Tian}
\affiliation{Institute of Space Physics, Luoyang Normal University, Luoyang, China}

%\author{Shuhong Yang}
%\affiliation{Key Laboratory of Solar Activity, National Astronomical
%Observatories, Chinese Academy of Sciences, Beijing 100012, China}

\begin{abstract}
%Taking advantage of the high temporal and spatial resolution observations from the {\sl Solar Dynamics Observatory},
In this paper, we report a detailed analysis of recurrent jets originated from a location with emerging, canceling and converging
negative magnetic field at the east edge of NOAA active region AR11166 from 2011 March 09 to 10. The event presented several interesting
features. First, a satellite sunspot appeared and collided with a pre-existing opposite polarity magnetic field and caused a
recurrent solar jet event. Second, the evolution of the jets showed blowout-like nature and standard characteristics. Third,
the satellite sunspot exhibited a motion toward southeast of AR11166 and merged with the emerging flux near the opposite polarity
sunspot penumbra, which afterward, due to flux convergence and cancellation episodes, caused recurrent jets. Fourth, three of the
blowout jets associated with coronal mass ejections (CMEs), were observed from field of view of the {\sl Solar Terrestrial Relations
Observatory}. Fifth, almost all the blowout jet eruptions were accompanied with flares or with more intense brightening in the jet base
region, while almost standard jets did not manifest such obvious feature during eruptions. The most important, the blowout jets were inclined
to faster and larger scale than the standard jets. The standard jets instead were inclined to relative longer-lasting. The obvious
shearing and twisting motions of the magnetic field may be interpreted as due to the shearing and twisting motions for a blowout jet eruption.
From the statistical results, about 30$\%$ blowout jets directly developed into CMEs. It suggests that the blowout jets and CMEs should have
a tight relationship.

\end{abstract}

\section{Introduction}

Solar jet activities are very common phenomena associated with magnetic flux emergence and
cancellation. They are observed both in cool and hot plasma ejections from the photosphere to the
outer corona in the solar atmosphere. Until to-date, many kinds of jet activities have been observed in
different spectral lines, such as H$\alpha$ surges \citep{roy73,liu04,jiang07}, extreme ultraviolet (EUV)
\citep{shen12,li15,liu15,shen11,nistico2015,kumar2017,zhu17,tian17,tianzhj2018} and
X-ray jets \citep{shib92,vasheghani2009,moor10,moor13,pucci2013,ster15,moor18}. Many researches indicated that most
solar jets result from magnetic reconnection in the low solar atmosphere \citep{shib92,shib94,can96,shimo96,shimo98,liuyu2002,jiang07,shen11,hong11,shen14a,li15,liu15,
ster15,li15,xuzhe16,hong17,shen17a,shen12,ad14,schmieder15,yang16,tianzhj2018,lixh2018,panesar2018,chengxin2018}.
In the past few decades, numerous characteristics of
the solar jets were found and summarized. According to \citet{shimo96,shimo98}, the authors found that jets have the
following features: 1) the general length is in the range of a few $\times$ 10$^{4}$ to 4$\times$ 10$^{5}$ km and the
width is 5$\times$10$^{3}$ to 10$^{5}$ km; 2) the lifetimes range from few minutes to more than ten hours;
3) the velocities are \speed{10} to \speed{1000}, and the average velocity is about \speed{200}; 4) 72$\%$
of them happened at the mixed polarity regions. \citet{liuyu2005} reported a detailed analysis of the relationship between
jet and jet-like CME. The authors indicated that a large scale jet probably can produced a jet-like CME.

By examining many X-ray jets, \citet{moor10} divided them into two types: one is standard jet and one is
non-standard jet, which is called blowout jet. In the classification, about two thirds of them are standard
type and about one third of them are blowout type. The authors also indicated that a blowout jet has enough
free energy to drive an ejective eruption, because the blowout jet has strongly shear or twist in the field
core of the arch. But, according to \citet{moor13}, the authors presented 54 polar X-ray jets. There were
32 blowout jets, 19 standard jets and 3 ambiguous jets, respectively. The number of the blowout jets has about twice as many as
standard jets. However, according to \citet{moor10}, the number of the standard jets twice over the blowout jets.
Combining the two samples, the authors indicated that the percentage of the blowout jets and standard jets is
closer to 50$\%$-50$\%$.
Some researches argued that jets could result in coronal mass ejections
(CMEs; \citealt{wang98,moore01,liu05,moore06,liu08,shen12,liu15,shen17a,sterlingreviw2018,tiwari2018}) and the
erupting plasmoid can be
observed in the corona \citep{moore07,liu08}. \citet{moor10} also constructed a model to interpret the blowout jet.
Blowout jets have tight association with small filament eruptions \citep{moor10,shen12,miao2018}, and they
usually accompany with flares during the eruptions. In the last few years, jet eruptions were reported to be associated
with CMEs by \citet{liu11}, \citet{shen12}, \citet{liu15}, \citet{miao2018} based on the high
spatial and temporal resolution of the {\sl Solar Dynamics Observatory} ({\sl SDO}; \citealt{pes12}).

Recurrent jets appear to often occurred in mixed polarity regions \citep{chen15,liujj2016}.
Magnetic reconnection is supposed to play a major role in the triggering mechanism of the jets
\citep{shib94}. Magnetic flux emergences, convergences and cancellations are the important features
in recurrent jet eruptions, that habitually appear near the pre-existing abient magnetic field \citep{jiang07,sterlingapj2018}.
In addition, many observational facts suggested that some jet activities can directly or indirectly result in
large-scale magnetic reconfigurations such as filament, loop eruptions \citep[e.g.,][]{jiang08,wang16,zheng16a,miao2018,shen18d},
CMEs, and EUV waves \citep{zheng12,zheng13,su15,2009ApJ...700L.182P,yanglh2017,shenyd2018b,shenyd2018c,miao2019}. It should be noted that \citet{shen12,miao2018}
%\textbf{\sout{respectively}}
presented a blowout jet eruption associated with simultaneous double-CME event. Blowout jet eruptions are considered as a
a very important solar phenomenon and usually generate narrow and bubble-like CMEs \citep{liu08,shen12,zheng13,archontis2013,
chen15,2017A&A...598A..41C,miao2018}. According to \citet{yashiro03} and \citet{vourlidas13}, a CME with angular width $\leq$ $20\degree$
is dubbed narrow CME.

In the present work we report an event occurred from 2011 March 09 11:00 UT to March 10 09:00 UT at the east periphery region
of AR11166 close to the solar disk center. The series jets included both
blowout-like and standard jets. Three CMEs were produced by
the counterpart of the three blowout jets. The paper is organized as follows. In Section 2, the reduction of the instruments and
data. In Section 3, the results are briefly listed. In the last section, the conclusions and discussions are presented.

\section{Instruments and data sets}

The recurrent jets were observed by the Atmospheric Imaging Assembly (AIA; \citealt{lemen12}) on board the {\sl SDO}
captures full-disk images of the Sun up to 1.3\rsun{} in seven EUV channels and three UV-visible channels, whose temporal
cadences are 12 s and 24 s, respectively, with a resolution of $0\arcsec.6$ pixel$^{-1}$. Six EUV wavelength channels:
94, 131, 171, 193, 211 and 335 \AA\ wavelengths are due to strong iron lines with temperature range from 0.6 MK to 16 MK.
The 304 \AA\ wavelength is due to \HeII{} line.
In this paper, only 304 and 171 \AA\ images are used to detect the jets.
The line-of-sight (LOS) magnetograms and continuum intensity images were taken by the Helioseismic and Magnetic Imager (HMI; \citealt{sche12})
on board the {\sl SDO}, whose temporal and spatial resolutions are 45 s and $0\arcsec.5$, respectively. The Space-weather HMI Active Region
Patches (SHARPs; \citealt{2014SoPh..289.3549B}) are also used for displaying the vector magnetic field evolution. The jets were also observed by the
Extreme Ultraviolet Imager (EUVI; \citealt{howard08}) of the Sun Earth Connection Coronal and Heliospheric Investigation
(SECCHI; \citealt{howard08}) on board the {\sl Solar TErrestrial Relations Observatory} ({\sl STEREO}; \citealt{kaiser08}) providing
full-disk He~{\sc ii} 304 \AA\ images, with a 5 and 10 minute cadences and a pixel width of $1\arcsec.6$. The three CMEs were
captured by the inner coronagraphs (COR1) and the outer coronagraphs (COR2) on board the {\sl STEREO}/Ahead (STA) and
{\sl STEREO}/Behind (STB), whose cadences and field of views (FOVs) are 5 minutes, 1.4--4\rsun{} and 2.5--15.6\rsun{} \citep{thomp03},
respectively. %The data from all the above mentioned instruments are used in our present study and analysis.

\section{Observations and Data Analysis}

\begin{figure*}[htbp]
\epsscale{1}
\plotone{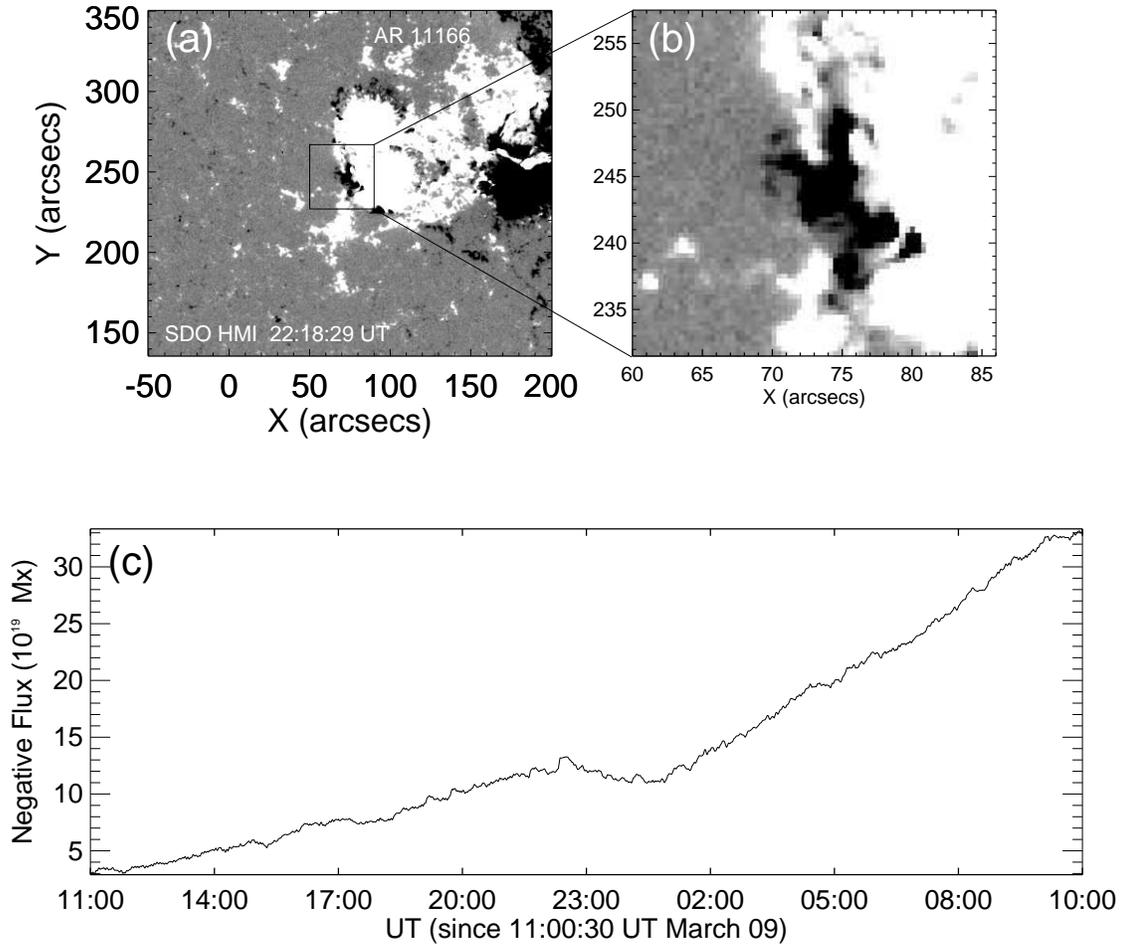}
\caption{Panel (a) is one HMI LOS magnetogram, with panel (b) consists of a zoomed view of the region marked by a black box in
panel (a), highlighting the position of the jet base negative flux. Panels (c) shows the box negative flux variation
from 11:00 UT on March 09 to 10:00 UT on March 10.
\label{hmi_flux}}
\end{figure*}

In defining and recognizing standard/blowout jets, \citet{moor10,moor13,moor15,moor18} assumed a relatively
simple magnetic structure, with a single bipole embedded in a background unipolar magnetic field. Based on that,
they defined ``standard'' and ``blowout'' jets based (in part) on the width of the spire compared to the size of the
jet's bipolar base, and based on observations in Hinode/XRT X-rays and AIA EUV images. Their setup was
representative of the magnetic situation for jets in coronal holes. The magnetic field at the base of our active
region jets is much more complicated than the simplified situation of Moore et al.. Therefore we cannot define our jets
as blowout or standard jets in exactly the same way that Moore et al. did. But, by looking at AIA EUV movies of the
jets of our region, we found many of them to tend to have relatively narrow spires compared to the width of the
negative patch shown in Figure \ref{hmi_flux}(b), and others of them to have spires that grew to be comparable in
width to that patch. Based on this, we classified the relatively-narrow-spire jets as standard jets, and the
relatively-broad-spire jets as blowout jets. We expect that our so-determined standard and blowout jets fit
closely with the categories defined by Moore et al..

Table \ref{tab:list} lists the blowout-like and standard jets from 2011 March 09 to 2011 March 10.
The table includes thirty-seven jets. Many of them are
standard jets (about 73$\%$ of the total statistical data), while the remaining events are blowout-like.
The jets are observed through 171 and 304 \AA\ wavelengths of the {\sl SDO}/AIA in the study. The
first column in the table displays the number of the jet events. The second and the third columns
in the table show the detection time of the jets. The fourth column displays the flares
associated with the jet eruptions. The fifth column reports the CMEs velocities from both COR2-A and COR2-B.
The sixth column presents the speeds of the jets and the seventh column shows the
duration time during the rise phase of the jets. The eighth column displays the lengths of the jets. We use
the $1\arcsec$ $ \approx $ 720 km\tablenotemark{\ref{long}}. The ninth column shows the types of the jets. The ``ST'' and
``BL'' denote the standard jet and the blowout jet, respectively. The characteristics of the standard jet
and blowout jet present several differences \citep{moor10,liu11,shen12,moor13,pucci2013,ster16,liujj2016}.
A movie highlighting all the jet eruptions, made using 171 and 304 \AA\ images (video1.mpeg) is given with the online
accompanying material. It should be noted that some very small jets are excluded because they appear to be too blurred
to be distinguished. The data at the beginning of the jet J11 is little blurred probably due to a little shake of the {\em SDO}.

\begin{figure*}[htbp]
\epsscale{1.2}
\plotone{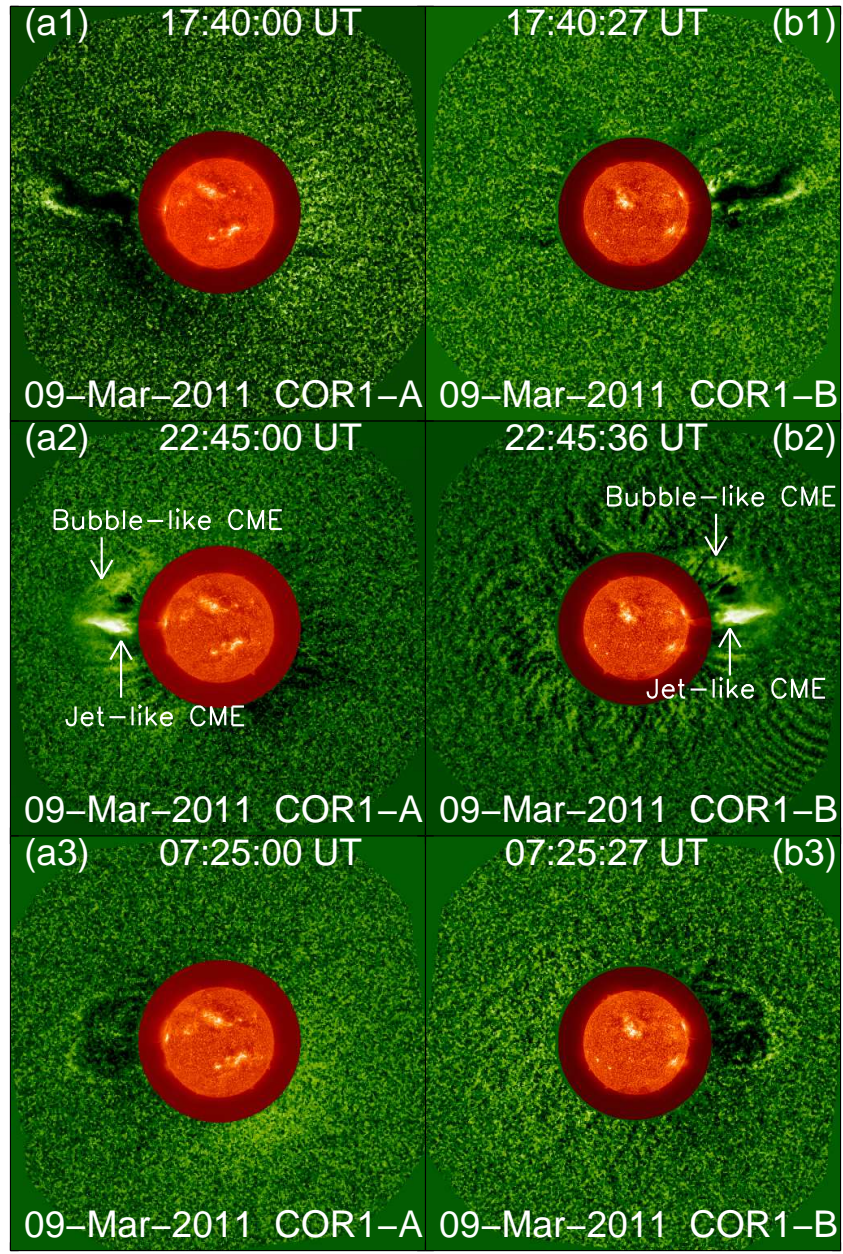}
\caption{Composite of {\sl STEREO} 304 \AA\ and running difference COR1 Ahead and Behind images showing the
three CMEs, which were triggered by J9, J15 and J35, respectively. Panels (a1), (b1) present the first CME event. Panels (a2), (b2)
show the double-CME event from the FOVs of {\sl STEREO} COR1 ahead and COR1 behind. The double-CME event has
two simultaneous CMEs, namely, the jet-like CME and the bubble-like CME, marked with the white
arrows. Panels (a3), (b3) display the third CME event.
\label{cme}}
\end{figure*}

Table \ref{tab:list2} lists the details of the three CME events, observed by {\sl STEREO}/COR2-A and COR2-B.
The three CMEs were detected on March 09 18:24 UT, 23:54 UT and March 10 07:54 UT from COR2-A, respectively.
On 2011 March 09 and 10, the separation angle between {\sl STEREO}/Ahead (STA) and {\sl STEREO}/Behind (STB) were
about $177^\circ$, while that between STA (STB) and {\em SDO} was about $88^\circ$ ($95^\circ$). Worth to note
that {\sl STEREO}/COR2-B data did not observe the third CME, while the
STEREO/COR1-B captured it (see Figure \ref{cme}(b3)). In the table, we focus on the COR2-A observed results
and list the COR2-B data only for reference. The CME events in Table \ref{tab:list2} can be found on the page SECCHI-A and -B CME
Lists\footnote{https://secchi.nrl.navy.mil/cactus/} from {\sl STEREO}. The seventh column displays the angular
width of the three CMEs. According to Table \ref{tab:list2}, two CMEs angular widths are $\leq$ 20$\degree$. The
two CMEs can be hence called narrow CMEs and they were triggered by the counterpart two blowout jet eruptions. Actually,
all of the three CMEs were triggered by blowout jets. Interestingly, \citet{miao2018} presented a double-CME event
associated with the jet J15. The authors displayed a simultaneous jet-like CME and bubble-like CME eruption that was
triggered by the jet J15 eruption. Hence strictly speaking, we observe four CMEs, three are
jet-like CMEs and a one bubble-like CME. According to the CME angular width, the CMEs can be classified as three
narrow CMEs and a bubble-like CME.

\begin{figure*}[htbp]
\epsscale{1.0}
\plotone{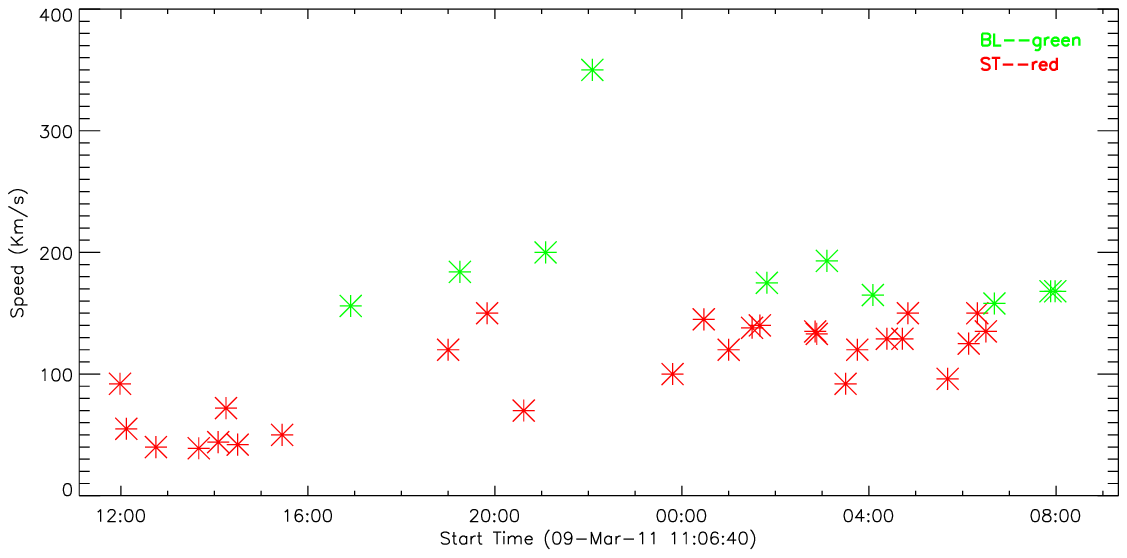}
\caption{The two type jets velocities evolutions. The red ``{\normalsize ${*}$}''
represents standard jets and the green ``{\normalsize ${*}$}'' represents blowout jets, respectively.
\label{jet_vel}}
\end{figure*}

Figure \ref{jet_vel} shows the speeds of the blowout and the standard jets. The red ``{\normalsize ${*}$}''
symbol represents the speeds of the standard jets and the green ``{\normalsize ${*}$}'' symbol represents the speeds
of the blowout jets', respectively. Combing with Table \ref{tab:list} and Figure \ref{jet_vel},
the blowout jets are faster and larger than the standard jets. According \citet{moor10}, a blowout
jet is more intense than a standard jet during the eruption stage. It should be noted that the three CMEs are
triggered by the blowout jets. On the one hand, this might explain why the blowout jet is actually more intense.
On the other hand, the blowout jet eruption tends to release large amounts of energy.

\begin{figure*}[htbp]
\epsscale{1.0}
\plotone{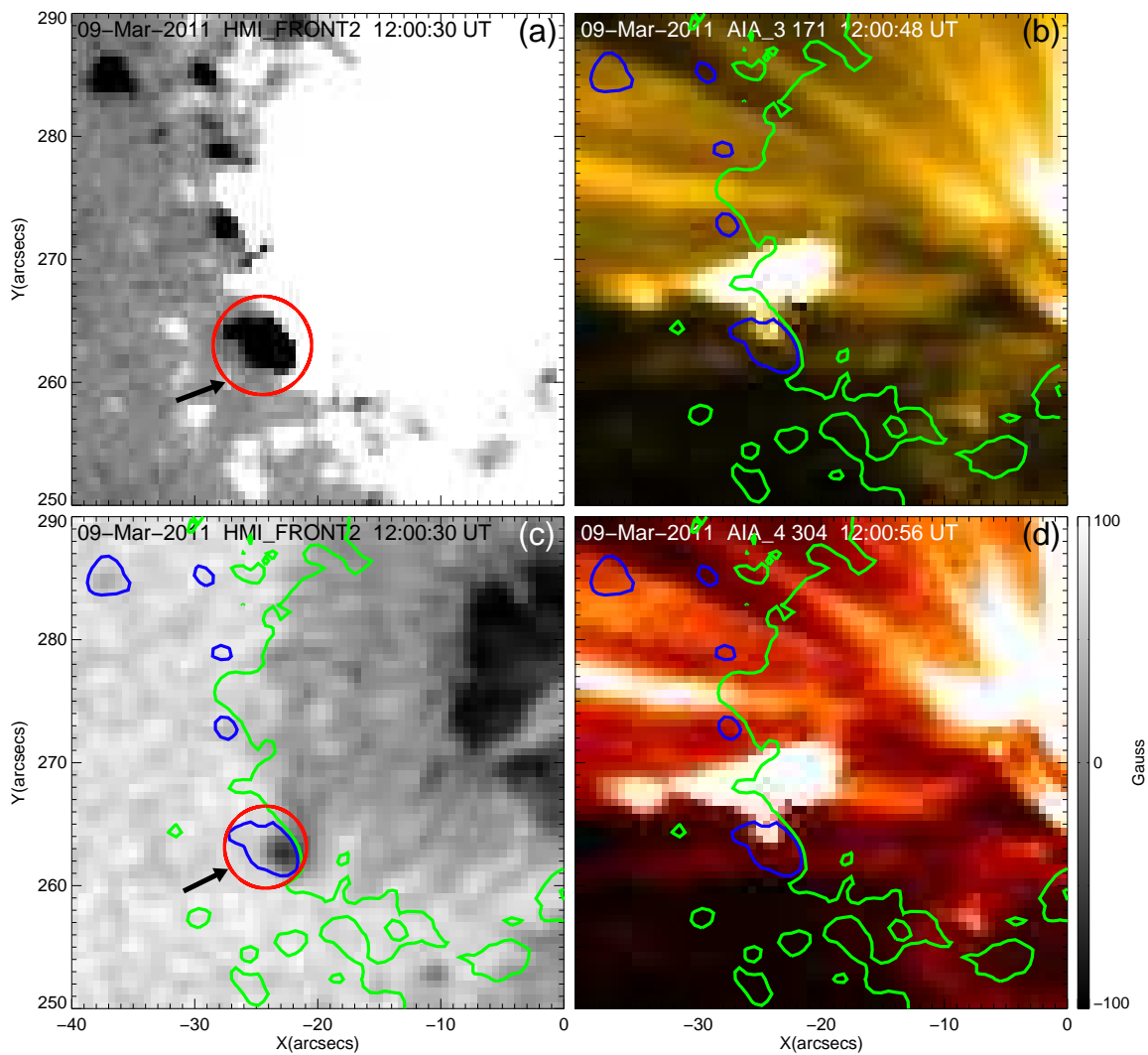}
\caption{Panels (a) and (c) display the HMI magnetogram and HMI continuum intensity image, respectively.
Panels (c), (b), and (d) show the contour levels $\pm100$ G. The blue and green lines represent negative
and positive flux,respectively. The red circles encompass regions highlighting the small sunspot in panel (a)
and the corresponding magnetic field in panel (c). Panels (b) and (d) show the first jet (J1) at
171 \AA\ and 304 \AA, respectively.
\label{sunspot}}
\end{figure*}

\begin{figure*}[htbp]
\epsscale{1.0}
\plotone{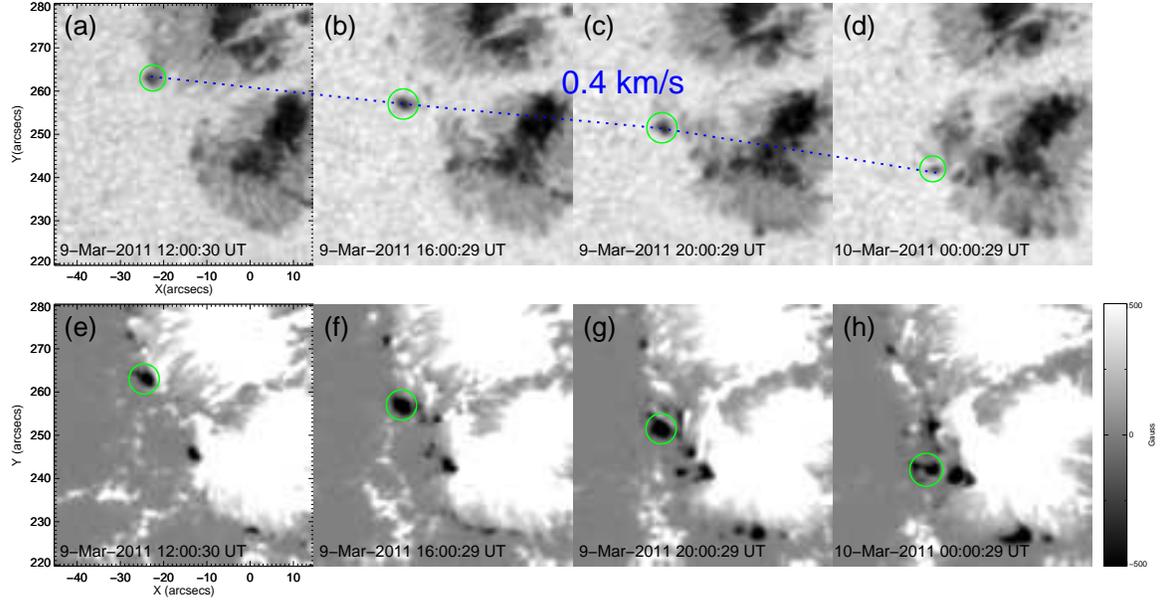}
\caption{A set of {\sl SDO}/HMI continuum intensity images displaying the satellite sunspot evolution and measured the moving average velocity
to be about \speed{0.4} in the top row. The bottom row shows the magnetic field evolution using
the {\sl SDO}/HMI magnetograms. The green circle regions represent the small sunspot and the corresponding  magnetic field
in the two rows, respectively. The blue dotted lines highlight the satellite sunspot motion path (see video2.mpeg).
\label{sunspot_vel}}
\end{figure*}

Interestingly, a satellite sunspot appeared near the main sunspot penumbra at about 11:30 UT.
In Figure \ref{sunspot} panel (b), at about 12:00 UT, the satellite sunspot was easily distinguishable at the east
edge of AR11166. Panel (a) shows the HMI LOS magnetic field of the small sunspot near the main pre-existing sunspot.
Panels (c) and (d) illustrate the first jet using the {\sl SDO}/AIA 171 and 304 \AA, respectively. ``J1'' is the first jet
in Table \ref{tab:list}. After 14:00 UT, the satellite sunspot was moving toward the southeast of the edge of AR11166 with
a very slow velocity. We measure the velocity of the satellite sunspot to be about \speed{0.4} (see the top row of
Figure \ref{sunspot_vel}). Panels (a) to (d) of Figure \ref{sunspot_vel} show the satellite sunspot evolution. Panels (e) to (h)
of Figure \ref{sunspot_vel} display the LOS magnetic field evolution from March 09 12:00 UT to March 10 00:00 UT.
%(video2 displays entire evolution stage of the magnetic field at the jet base region).
Later on, the magnetic field of the jet base gets more and more
complex and the negative magnetic field became more intense.
In order to display the satellite sunspot and the counterpart of magnetic field evolutions, we provide a movie
(video2.mpeg; in the online journal), constructed using HMI continuum intensity images and LOS magnetograms.

\begin{figure*}[htbp]
\epsscale{1.2}
\plotone{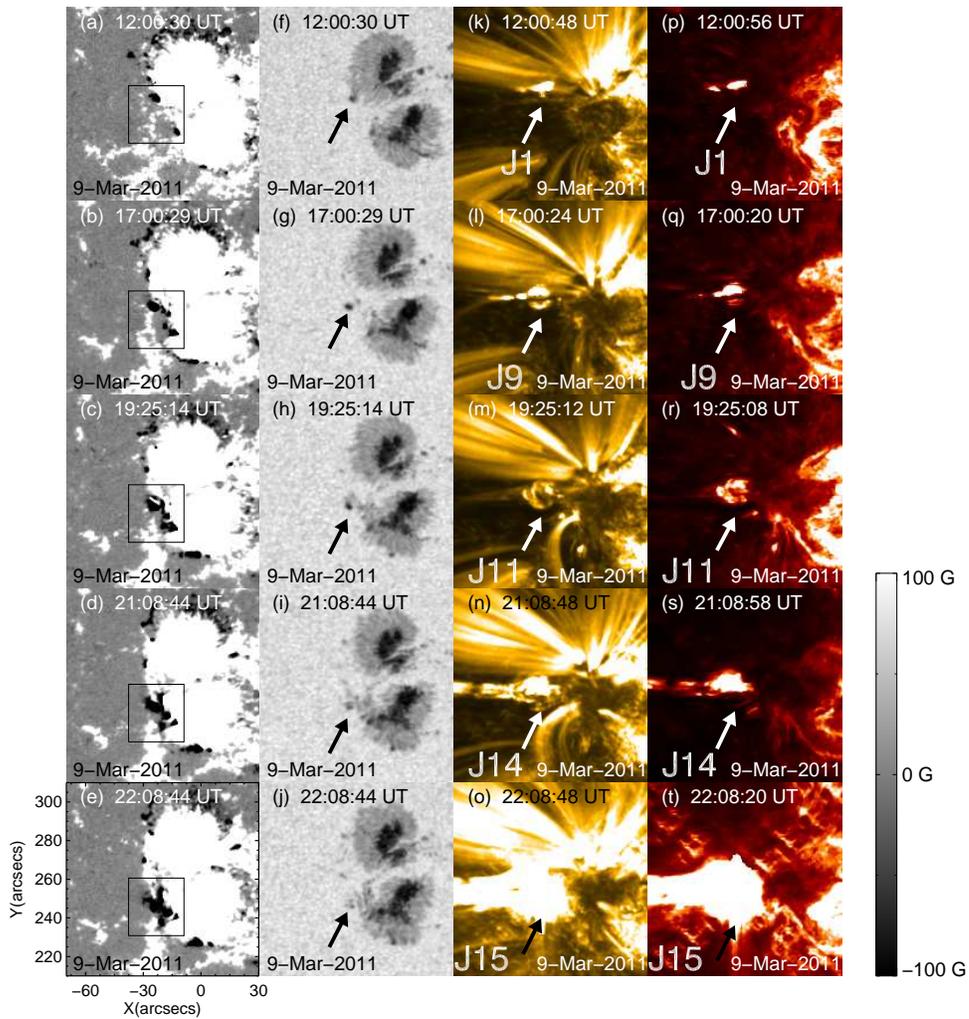}
\caption{The first column and the second column show the magnetic field and the satellite sunspot evolutions, respectively.
The third and fourth column correspond to jet eruptions at 171 \AA\ and 304 \AA. The five jets occurred on 09 March 2011.
The more jets are shown in the online material (see video1.mpeg in the online journal).
\label{jets}}
\end{figure*}

Five jets are displayed in Figure \ref{jets}. Panels (a) to (e) in the first column correspond to HMI LOS magnetograms
highlighting evolution of the jet base region magnetic flux. Panels (f) to (j) in the second column are HMI continuum
intensity images displaying the evolution of the satellite sunspot. The third and fourth columns are 171 and
304 \AA\ images for the five jets. They counterpart with the magnetic flux evolution of the jet base region,
which we report in the first and second columns. the other jets can be recognized in the online animation material: video1.mpeg.
From the first and the second columns, the magnetic field evolution is very clear. Worth to mention here that we mainly focus on the
negative flux evolution, because the pre-existing sunspot is too big to accurately measure the positive flux evolution.

In order to inspect the negative flux evolution of the jet base region, we measure data from March
09 11:00:30 UT to March 10 10:00:30 UT.
%  flux Figure \ref{hmi_flux}, the negative flux was measured from March 09 11:00:30 UT to March 10 10:00:30 UT.
The results are displayed in Figure \ref{hmi_flux}. Panel (a) shows AR11166 with the box indicating the
jet base region. The box region is shown in panel (b) of Figure \ref{hmi_flux}. Panel (c)
presents the evolution of the negative flux at jet base region. It is so important to pay more attention to the negative
flux region. The positive flux region is too large and complex which impedes a precise quantification of the positive flux.
The result suggests that the negative flux shows manifests both emergence and cancellation episodes during the whole stage.
The overall trend indicates emergence to be more intense and faster than cancellation. It should be noted that the
cancellation is undergoing on both the north and south sides of the negative patch, which near the pre-existing positive polarity.
Actually, the jets come from both of these locations. \citet{ster17} indicated that eruptions of strong jets occur only during times
of intense flux cancellation. Our observations are consistent with this, because our jets come from the two locations where
we see cancellation.

%Due to the jet region
%is too closed the main magnetic field to accurate measure the positive flux, we only focus on the negative flux.

%It should be noted that during the J15 eruption stage from March 09 22:00 UT to March 10 01:00 UT.

According to \citet{yanglh13} and \citet{chen15}, the jets are usually produced by the magnetic reconnection or cancellation.
In order to understand the evolution of the satellite sunspot and magnetic field, we analyze vector magnetograms around the
%panels (a) to (d), we analyzed the vector magnetograms using the same measure in \citet{xu16}.
jet region. In Figure \ref{vector_mag}, the roman numerals ``I'', ``II'', ``III'', represent the satellite sunspot flux region,
emergence flux region and the whole of the jet base region, respectively. To distinctly scrutinize the vector magnetic
flux evolution, we present a movie (video3.mpeg) of the evolution of the magnetic field of the jet base region.
From video3.mpeg, we could easily observe the evolution of the vector magnetic flux in the jet base region. The
satellite sunspot moved towards the southeast of the pre-existing sunspot and merged with the emergence flux from March 09
to March 10. The video3.mpeg also displays characteristics of strong twisting motion and shearing motion of the
magnetic field at the jet base region.

During the time interval March 09 11:00 UT to March 10 08:00 UT, there were three large scale CMEs produced by blowout jet eruption.
The details of the three CMEs are listed in Table \ref{tab:list} and Table \ref{tab:list2}, which are observed from the {\sl STEREO}/COR2.
In Figure \ref{cme} we display the three CMEs. Snapshots are 304 \AA\ and running difference
COR1 images at the limb as viewed from {\sl STEREO} Ahead and Behind.

Figure \ref{cme} reports the composite EUVI 304 \AA\ and COR1 images. The structures of the three CMEs
can be clearly seen. Panels (a1), (b1) show the first CME at about 17:40 UT. Panels (a2), (b2) display the second CME, which is a
double-CME event (see \citet{miao2018} for more details). Panels (a3), (b3) present the third CME. The three CMEs have a common
characteristic namely being triggered by the blowout jets.

%The second CME is associated with a mini-filament eruption
%due to the {\sl STEREO} data website has something wrong, some data can not download.

%The second CME is associated with double CMEs. The blowout jet
%has a large scale, according to \citet{miao18}. From the FOVs of {\sl STEREO}/EUVI, COR1 and COR2 Ahead, we present two movies
%(video4\footnote{https://helioviewer.org\label{web}}, video5\textsuperscript{\ref{web}}). Figure \ref{cme} shows the simultaneous
%double-CME event from two FOV of {\sl STEREO}.

\begin{table*}[h]\tiny%\footnotesize%\normalsize
\begin{center}
\setlength{\tabcolsep}{2.5mm}{
\caption{Date and time for the observed jets, and their measured parameters. \label{tab:list}}
%\begin{tabular}{crrrrrrrrr}
\begin{tabular}{c*{9}{c}}
  \noalign{\smallskip}\tableline\tableline \noalign{\smallskip}
  \multicolumn{9}{l}
 {\textbf {Jets:}} \\
    \noalign{\smallskip}\tableline \noalign{\smallskip}
  Jet No. & Date & Time &  Flare &  CME Speed
   &  Jet Speed & Jet Rise Dur. & Jet Length\footnote{We use the 1 arcsec $ \approx $ 720km \label{long}} & Type \\  %\tablenotemark{e}
  & & (UT)  & class & (\kms)  & (\kms) & ($\pm$ 1 min) & ($\pm$ 1500 km)& ST or BL\footnote{BL means Blowout and ST means Standard, respectively.} \\

\noalign{\smallskip}\tableline \noalign{\smallskip}

J1 & 09-Mar-11  & 11:59:08  & \nodata & \nodata & 92 & 13 & 72000 & ST \\
J2 & 09-Mar-11  & 12:07:32  & \nodata & \nodata & 55 & 13 & 43200 & ST\\
J3 & 09-Mar-11  & 12:45:56  & \nodata & \nodata & 40  & 15 & 36000 & ST \\
J4 & 09-Mar-11  & 13:40:32  & \nodata & \nodata & 39 & 20 & 46700 & ST  \\
J5 & 09-Mar-11  & 14:05:08  & \nodata & \nodata & 44 & 15 & 39600 &  ST \\
J6 & 09-Mar-11  & 14:15:20  & \nodata & \nodata & 72 & 10 & 43200 &  ST \\
J7 & 09-Mar-11  & 14:30:20  & \nodata & \nodata & 42 & 17 & 43200 &  ST \\
J8 & 09-Mar-11  & 15:27:20  & \nodata & \nodata & 50 & 18 & 50400 &  ST\\
J9 & 09-Mar-11  & 16:55:32  & \nodata &    256\footnote{https://secchi.nrl.navy.mil/cactus/index.php?p=SECCHI-A/2011/03/out/Flow0005/CME.html\label{cme1a}},
126\footnote{https://secchi.nrl.navy.mil/cactus/index.php?p=SECCHI-B/2011/03/out/Flow0009/CME.html\label{cme1b}}
& 156 & 23 & 216000 &  BL \\

J10 & 09-Mar-11  & 19:00:20 & \nodata & \nodata & 120 & 15 & 108000 &  ST  \\
J11 & 09-Mar-11  & 19:15:53 & \nodata & \nodata & 184 & 13 & 144000 & BL \\
J12 & 09-Mar-11  & 19:50:08 & \nodata & \nodata & 135 & 16 & 129600 & ST\\
J13 & 09-Mar-11  & 20:37:32 & \nodata & \nodata & 70 & 26 & 108000 & ST\\
J14 & 09-Mar-11  & 21:05:08 & \nodata & \nodata & 200 & 18 & 216000 & BL\\
J15 & 09-Mar-11  & 22:05:08 & C9.4\footnote{https://www.solarmonitor.org/?date=20110309 \label{flare}}
  & 337\footnote{https://secchi.nrl.navy.mil/cactus/index.php?p=SECCHI-A/2011/03/out/CME0032/CME.html \label{cme2a}},
  143\footnote{https://secchi.nrl.navy.mil/cactus/index.php?p=SECCHI-B/2011/03/out/CME0030/CME.html\label{cme2b}}
  &  350\footnote{See the velocity in the paper \citet{miao2018}}  & 25 & 435000 & BL  \\
J16 & 09-Mar-11  & 23:48:20  & \nodata & \nodata & 100  & 12 & 72000 & ST \\
J17 & 10-Mar-11  & 00:28:32  & \nodata & \nodata & 145 & 12 & 104400 & ST \\
J18 & 10-Mar-11  & 01:00:20  & \nodata & \nodata & 120  & 15 & 108000 & ST \\
J19 & 10-Mar-11  & 01:30:20  & \nodata & \nodata & 138  & 10 & 108000 & ST \\
J20 & 10-Mar-11  & 01:40:32  & \nodata & \nodata & 140  & 9 & 75600 & ST \\
J21 & 10-Mar-11  & 01:49:32  & \nodata & \nodata & 175 & 12 & 126000 & BL \\
J22 & 10-Mar-11  & 02:51:20  & \nodata & \nodata & 135 & 12 &  97200 & ST \\
J23 & 10-Mar-11  & 02:53:08  & \nodata & \nodata & 133 & 11 & 72000 & ST \\
J24 & 10-Mar-11  & 03:06:56  & \nodata & \nodata & 193 & 13 & 151200 & BL \\
J25 & 10-Mar-11  & 03:30:20  & \nodata & \nodata & 92 & 13 & 72000 & ST \\
J26 & 10-Mar-11  & 03:45:20  & \nodata & \nodata & 120 & 10 & 72000 & ST \\
J27 & 10-Mar-11  & 04:05:08  & C2.9\footnote{https://www.solarmonitor.org/?date=20110310}
 & -     & 165 & 12 & 118000 & BL \\
J28 & 10-Mar-11  & 04:23:08  & \nodata & \nodata & 129 & 15 & 116000 & ST \\
J29 & 10-Mar-11  & 04:43:32  & \nodata & \nodata & 129 & 15 & 116000 & ST \\
J30 & 10-Mar-11  & 04:50:08  & \nodata & \nodata & 150 & 10 & 90000 & ST \\
J31 & 10-Mar-11  & 05:41:08  & \nodata & \nodata & 96 & 10 & 57600 & ST \\
J32 & 10-Mar-11  & 06:08:08  & \nodata & \nodata & 125 & 13 & 97200 & ST \\
J33 & 10-Mar-11  & 06:19:32  & \nodata & \nodata & 150 & 12 & 108000 & ST \\
J34 & 10-Mar-11  & 06:30:56  & \nodata & \nodata & 135 & 8 & 64800 & ST \\
J35 & 10-Mar-11  & 06:41:08  & \nodata &  263\footnote{https://secchi.nrl.navy.mil/cactus/index.php?p=SECCHI-A/2011/03/out/CME0034/CME.html \label{cme3a}}
 & 158 & 19 & 180000 & BL \\
J36 & 10-Mar-11  & 07:53:08  & \nodata & \nodata & 168 & 5 & 50400 & BL \\
J37 & 10-Mar-11  & 07:59:08  & \nodata & \nodata & 168 & 5 & 50400 & BL \\

\tableline
\end{tabular}}

%\footnotetext{https://secchi.nrl.navy.mil/cactus/index.php?p=SECCHI-A/2011/03/out/Flow0005/CME.html} %256\tablenotemark{c}
  %256
%\tablenotemark{g}                                           %C9.4
   %337
%\footnotetext{The data comes from \citep{miao18}}
 %263
%\footnote{ftp://ftp.ngdc.noaa.gov/STP/space-weather/solar-data/solar-features/solar-flares/x-rays/goes/2014/}
%\footnotetext{http://cdaw.gsfc.nasa.gov/CME$\_$list}
%\footnotetext{The uncertainty in the CMEs speed measurement is less than 10$\%$ \citep{yashiro04}.}
%\footnotetext{The uncertainties are estimated from the time-distance plots.}
%\footnotetext{Measured at a projected height of $\sim$72000 km from jet base.}
%\footnotetext{This jet shows up well in the AIA 94 \AA\ images, but not in 304 \AA\ images. Due to its poor visibility
%in 304 \AA\ images, we were unable to follow the jet plasma well enough to measure its speed.}%%
%The uncertainty in the jet speed and width measurements are  $\pm$ 20 \kms and $\pm$ 1500 km respectively.}

%\textsuperscript{\ref{web}}
%video7\footnote{https://cor1.gsfc.nasa.gov/catalog/cme/2011/html/201103092145a.html}
%and
%video8\footnote{https://cor1.gsfc.nasa.gov/catalog/cme/2011/html/201103092145b.html};
%(see the COR2 two views online materials on March 10, 2011\footnote{https://cdaw.gsfc.nasa.gov/stereo/}).

% as well as readers can easily find movies on 09 March 2011 in the CME list daily movies

%ORL\footnote{https://blog.csdn.net/bear\_kai\label{web}} and
%Jaffe\textsuperscript{\ref {web}}

\end{center}
\end{table*}

\begin{table*}[h]\tiny %\footnotesize  %\normalsize
\begin{center}
\setlength{\tabcolsep}{2.5mm}{
\caption{Date and time for the observed jets, and their measured parameters. \label{tab:list2}}
%\begin{tabular}{crrrrrrrrr}
\begin{tabular}{c*{9}{c}}
  \noalign{\smallskip}\tableline\tableline \noalign{\smallskip}
  \multicolumn{9}{l}
 {\textbf {CME-producing Jets:}} \\
    \noalign{\smallskip}\tableline \noalign{\smallskip}
   Jet No. &  CME No. & Date & Time &  Flare &  CME Speed & CME Angular & CME Rise Dur. & Jet Type & \\  %\tablenotemark{e}
  & STEREO/COR2-A &  & (UT)& Class  &(\kms)  & Width (\degree) & (hour) & &\\

  J9 & cme1a\tablenotemark{\ref{cme1a}} & 09-Mar-11 & 18:24 & \nodata &  256 &  20  & 01 & BL & \\

  J15 & cme2a\tablenotemark{\ref{cme2a}}& 09-Mar-11 & 23:54 & C9.4\tablenotemark{\ref{flare}} & 337 & 44 & 03 & BL & \\

  J35 & cme3a\tablenotemark{\ref{cme3a}} & 10-Mar-11 & 07:54 & \nodata & 263 & 14  & 01 &  BL & \\

\noalign{\smallskip}\tableline \noalign{\smallskip}

%\noalign{\smallskip}\tableline \noalign{\smallskip}\\

Jet No. &  CME No. & Date & Time &  Flare &  CME Speed & CME Angular & CME Rise Dur. & Jet Type & \\  %\tablenotemark{e}
  & STEREO/COR2-B  &   & (UT)& Class  &(\kms)  & Width (\degree) & (hour) & &\\

  J9 & cme1b\tablenotemark{\ref{cme1b}} & 09-Mar-11 & 18:24 & \nodata&  126 &  16  & 00 & BL & \\

  J15 & cme2b\tablenotemark{\ref{cme2b}} & 09-Mar-11 & 23:24 & C9.4\tablenotemark{\ref{flare}} & 143 & 24 & 01 & BL & \\

  J35 & \nodata & 10-Mar-11 & \nodata & \nodata &\nodata &  \nodata & \nodata &  BL & \\

\noalign{\smallskip}\tableline \noalign{\smallskip}

\tableline

\tableline
\end{tabular}}
\end{center}
\end{table*}

\section{Conclusions and Discussions}
 The solar recurrent jets occurring from 2011 Mach 09 12:00 UT to 2011 March 10 08:00 UT have been investigated in the present work.
 The jets can be classified into two types, namely, blowout jet and standard jet. We mainly focus on
 the recurrent jets and three CMEs associated with blowout jets. The main observational results can
 be summarized as follows.

 1. The recurrent jet eruption associated with a newly formed satellite sunspot, together with a negative flux appeared in the
 penumbra of the main sunspot at about 11:30 UT on March 09. After 14:00 UT, the penumbra southeast of AR11166,
 presents a scenario whereby the negative pole of an emerging-flux element is growing in a sea of positive polarity.
 We measure the speed of the small satellite sunspot to be \speed{0.4}.

 2. During the whole process, ten blowout jets and twenty-seven
 standard jets were observed. About 27$\%$ are the blowout jets while 73$\%$  belong to standard jets.

 3. All the blowout jet velocities are found to be larger than those of standard jets. We report the velocity results
 in Table \ref{tab:list}.
 Compared to standard jets, blowout jets are generally brighter and more explosive in EUV, suggesting that they
 are undergoing intense magnetic reconnection. This may suggest that the blowout jet releases more energy than standard jets.

 4. The eruption processes of magnetic cancellation and emergence are clearly identified and reported. From the video2.mpeg and
 the video3.mpeg, we can evidently recognize the magnetic field evolution at the the jet base region. Many new small emerging
 fluxes collided and converged during the whole stage. Especially, the obvious shearing and twisting motions of the
 magnetic field are probably interpreted as indication of the shearing and twisting motions for blowout jet eruptions. When a blowout jet
 erupted, the shearing and twisting motions of the magnetic field (see the video2.mpeg and video3.mpeg) appear more intense than a standard jet.

 5. The three CMEs were produced by the corresponding blowout jet eruptions. All of the jets are displayed in Table \ref{tab:list},
 where the jet J15 has the largest scale. This blowout jet led to a double-CME event \citep{miao2018}. Blowout jets were easily observed in the
 region of the recurrent jets, with distinct complex magnetic field and mixed polarities. From the statistical results, about 30$\%$ blowout
 jets directly developed into CMEs. Due to the low resolution of the {\em STEREO}, we consider that is probably more than 30$\%$ blowout jets
 can directly developed into CMEs. The forecast requires of course more statistical data to be confirmed in the future.

In Table \ref{tab:list}, ten blowout jet eruptions and twenty-seven standard jet eruptions were observed from 2011 March 09 to March 10.
According to \citet{moor10}, the authors observed the interior base arch brightened with the production of the blowout jet. They
thought that in blowout jet the interior of the base arch takes part in the eruption. According to the video1.mpeg, all of
the blowout jets brightened in the interior of the base arch with the blowout jet eruptions. The brightening in the interior
of the base arch becomes more extensive. In Table \ref{tab:list} and Table \ref{tab:list2}, two large blowout jet eruptions
are associated with C class flares.

\begin{figure*}[htbp]
\epsscale{1.2}
\plotone{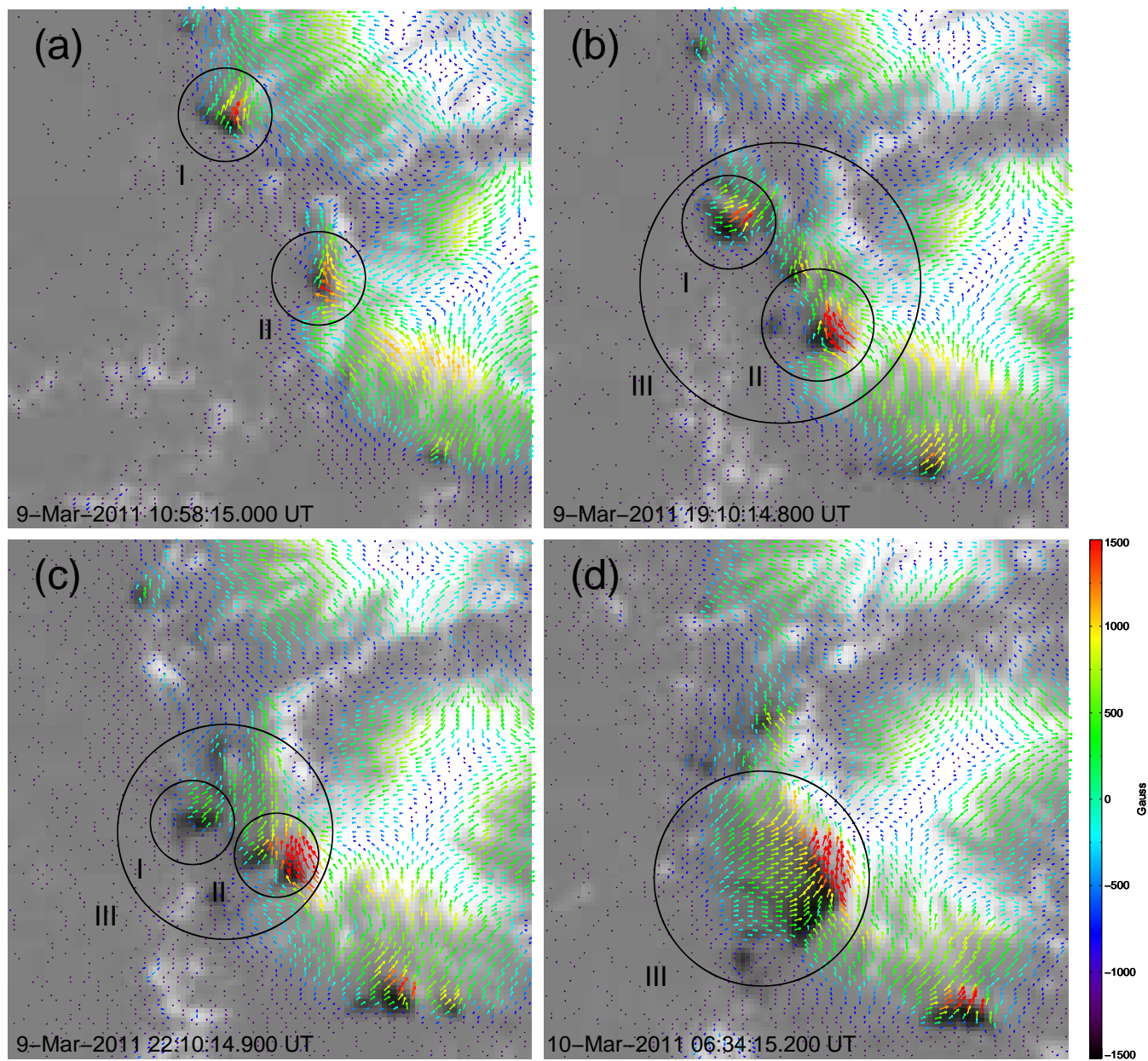}
\caption{Panels (a)-(d) display the vector magnetic evolution. The Roman numerals ``I'' and ``II'' represent satellite sunspot magnetic field and magnetic
emergence field, respectively. The ``III'' displays the mixed and convergence magnetic field region (see video3.mpeg).
\label{vector_mag}}
\end{figure*}

The magnetic emergence and cancellation both play an important role in this jet activities. There are few events associated with
a new sunspot triggering the recurrent jet events. Interestingly, the new satellite sunspot appears to move toward the emergence
region in the southeast of the main sunspot. The satellite sunspot had negative flux as the same emergence flux. At about 21:00 UT
on March 09, the satellite sunspot and the emergence flux converged together. From Figure \ref{vector_mag}, during the evolution
of the vector magnetic, the jet base region had intense cancellation (see video2.mpeg; material online). It suggests that the jet base
region embodies very intense flux emergence feature. The negative flux evolution indicates that the emergence, cancellation and convergence
co-exist at the same time. Through all the study stage, the emergence appears more intense than the cancellation.

According to \citet{moor10,wang10,chen15}, the eruption of blowout jet requires a large amount of energy. From the
video1.mpeg and Table \ref{tab:list}, we can see that the blowout jet eruptions accompany the brightening in the
jet base region at the beginning. The jet J15 eruption accompanied with the C9.4 flare was observed at about 22:05 UT.
In Table \ref{tab:list}, the scale of most of blowout jets are larger than standard jets and the velocities are
faster than standard jets \textbf{\citep{pucci2013}}.
The results indicate that a blowout jet eruption need more energy than a standard jet. \textbf{\citet{liujj2016,pucci2013}} reported that
during the magnetic reconnection of blowout jet more free energy should be released compared to standard jets. It suggested that
the scale of a blowout jet should be usually larger than a standard jet. According to
Figure \ref{vector_mag} and video3.mpeg, magnetic emergence and cancellation were clearly very intense. We believe that cancellation
has provide enough amount of energy to keep the recurrent jet eruptions. Even some blowout jet eruptions can produce CMEs, such as jet J9, J15
and J35 in the listed in Table \ref{tab:list} and Table \ref{tab:list2}. \citet{liu08} has studied the relationship between
surges and CMEs. Due to the low cadence of the data in \citet{liu08}, we considered that the jet-like surges probably are blowout jets.
The narrow CMEs were formed directly from the blowout jet eruptions. In addition and according to \citet{liu08,moor10} and Table \ref{tab:list},
\ref{tab:list2}, The statistical results support that the blowout jets and CMEs have a tight relationship. Our results
therefore might suggest that a blowout jet has larger scale and needs more free magnetic reconnection energy than a standard
jet. More similar observational investigations will be reported to confirm and support our present findings in the future.

In this paper, we reported the finding of thirty-seven jets and three CMEs as displayed in Table \ref{tab:list}
and Table \ref{tab:list2}. Our study suggests that the blowout or blowout-like jets are larger and faster than the standard jets.
Another intriguing result indicates that the blowout type eruptions can more easily trigger CMEs than standard type. The blowout
jet eruptions tend to be accompanied with more intense flares or brightenings. We believe many new characteristics can be reported from
recurrent jet eruptions in the future, which may aide to promote hence the development of new physical aspects of these interesting events.

%In this paper, many jets were investigated, but as the region was very active. Many small indistinct jets can not be
%distinguished and some jets bursted very fast blended with other jets. As the above reasons, may be some jets
%can not exactly record in Table \ref{tab:list}. However, as the number of homologous jets studied in the paper is limited.
%The blowout jets and standard jets percentage nearly the data in the \citet{moor10}. In the future, we will study statistical
%analysis of more homologous jets from different active regions, as well as include statistics of the standard jets and blowout
%jets.

\acknowledgments We are grateful to the {\em SDO} and {\em STEREO} teams for the excellent data they provided.
 We also thank the referee for valuable suggestions and comments that improved the quality of this paper.
 This work is funded by the grants from the Strategic Priority Research Program of CAS with grant XDA-17040507,
 the National Scientific Foundation of China (NSFC 11533009, 11773068) and from the Key Laboratory of Geospace
 Environment, CAS, University of Science and Technology of China. This work is also supported by the grant
 associated with project of the Group for Innovation of Yunnan Province. The authors Y. Liu and Z.Z. Abidin
 would like to thank the University of Malaya Faculty of Science grant (GPF040B-2018) for their support. The
 authors extend their appreciation to the Deanship of Scientific Research at King Saud University for funding
 this work through research group No. (RG-1440-092). In addition, we are also grateful to the One Belt and One
 Road project of the West Light Foundation, CAS.

%\bibliography{myhrefs}
%\bibliographystyle{aasjournal}

\end{document}